# Ultrafast THz Faraday Rotation in Graphene

J. N. Heyman[1], R. F. Foo Kune[1], B. A. Alebachew[1], M. D. Nguyen[1], J.T. Robinson[2]

1. Macalester College, St. Paul, MN 55105
2. Naval Research Laboratory, Washington DC 20375

**ABSTRACT**

Terahertz (THz) Faraday rotation measurements were performed to investigate carrier dynamics in *p*-type CVD graphene. We used static and time-resolved polarization-sensitive THz transmission measurements in a magnetic field to probe free carriers in GaAs, InP and Graphene. Static measurements probe the equilibrium carrier density and momentum scattering rate. Time-resolved (optical pump/THz probe) measurements probe the change in these quantities following photoexcitation. In a typical CVD graphene sample we found that 0.5ps following photoexcitation with $1\cdot10^{13}$ photons/cm$^2$ pulses at 800nm the effective hole scattering time decreased from 37fs to 34.5fs, while the carrier concentration increased from $2.0\cdot10^{12}$cm$^{-2}$ to $2.04\cdot10^{12}$cm$^{-2}$, leading to a transient decrease in the conductivity of the film.

I.    Introduction

Graphene is a single-atom thick layer of carbon arranged in a honeycomb lattice with a gapless band structure and linear bands near the Fermi energy. This 2D material exhibits extraordinary mechanical, electronic and optical properties and is a prime candidate for electronic and optoelectronic applications. Chemical vapor deposition (CVD) has emerged as a practical technique for growing large-area graphene films that can be transferred to any substrate. These transferred CVD-graphene films typically have higher carrier densities and lower mobilities compared to exfoliated graphene. Understanding the dynamics of free carriers in this CVD material and their response to optical excitation is important for possible applications.

A significant amount of recent research has used time-resolved probes to study hot carrier dynamics in graphene. In these studies, the material is excited with an optical or infrared pump pulse and the system is probed with photoelectron[1-4] or optical spectroscopy at visible[5-9] or infrared[10, 11] frequencies. The majority of these studies are consistent with a single model: Carrier-carrier and carrier-phonon intraband scattering produces distinct Fermi distributions of electrons and holes ~50fs following photoexcitation. Interband scattering merges these distributions, producing a thermal distribution of carriers in the valance and conduction band with a common chemical potential within ~300fs. Carrier-phonon scattering causes the hot electrons and holes to equilibrate with the optical phonons within ~1ps. The ensemble of hot carriers and phonons cools to equilibrium as the optical phonons decay over longer timescales.

Time-resolved THz spectroscopy[12-22] has also been used to study free carriers in graphene. Graphene typically exhibits strong free carrier absorption at THz frequencies that can be fit by the Drude model to yield the carrier density and effective scattering rate. The interband absorbance in doped graphene also shows a low-frequency cut-off at $\hbar\omega \approx 2\varepsilon_F$ (typically in the mid-infrared) that can be used to independently determine the free carrier density[23]. Time-resolved THz (Optical pump/THz probe) spectroscopy can study free carrier dynamics in graphene following photoexcitation by determining the change in the frequency-dependent conductivity.

Time-resolved THz studies of graphene have reported both transient increases and decreases in conductivity following photoexcitation. Measurements of gated samples show [21] that the sign of the response is controlled by carrier density, with highly conductive samples showing negative photoconductivity (*increase* in THz transmission) and low carrier density samples showing positive photoconductivity. Two competing effects are believed to determine the sign of the conductivity change: photoexcitation increases the free carrier density which increases conductivity, but carrier heating increases the scattering rate due to electron-phonon scattering, decreasing it.

To effectively probe this phenomenon, we independently probe carrier density and carrier scattering rate following photoexcitation. In principle, optical pump/THz probe measurements of the frequency dependent conductivity can independently resolve changes in the carrier density and scattering rate. However, experimental artifacts can easily mask or distort the conductivity spectrum, making such analysis unreliable, particularly in samples with carrier scattering rates much larger than the maximum frequency probed.

In this study we use polarization-sensitive THz transmission measurements in a magnetic field to independently determine the conductivity and the carrier scattering rate, and we apply this technique to probe carrier dynamics in graphene. This technique allows us to determine the carrier scattering rate without fitting the frequency-dependent conductivity to the Drude model.

The technique we have developed is similar to the THz Optical-Hall Effect technique developed by Hoffman [24]. For small, non-quantizing fields ($\hbar/\tau \gg \Delta\varepsilon_{L.L.}, |\omega_c\tau| \ll 1$) the conductivity tensor of a 2D sheet in a perpendicular magnetic field is:

$$\begin{pmatrix} \sigma_{xx} & \sigma_{xy} \\ \sigma_{yx} & \sigma_{yy} \end{pmatrix} = \frac{\sigma_0}{1+\omega_c^2\tau^2} \begin{pmatrix} 1 & \omega_c\tau \\ -\omega_c\tau & 1 \end{pmatrix}, \qquad (1)$$

where $\sigma_0$ is the zero-field conductivity, $\omega_c$ is the orbital cyclotron frequency, $\tau$ is the carrier scattering rate and $\Delta\varepsilon_{L.L.}$ is the Landau-Level spacing at the Fermi Energy. The ratio $\sigma_{xy}/\sigma_{xx} = \omega_c\tau$ is proportional to the magnetic field and determines the carrier scattering time.

The conductivity tensor components $\sigma_{xx}$ and $\sigma_{xy}$ can be determined from polarization-sensitive THz transmission measurements. Consider a linear polarized electromagnetic wave normally incident on a thin conducting sheet on a dielectric substrate and suppose the electric field amplitude and polarization of the transmitted wave can be measured. Let the parallel transmittance be defined $T_\parallel(\omega) = E_\parallel(\omega)/E_\parallel^0(\omega)$, where $E_\parallel(\omega)$ is the component of the transmitted electric field parallel to the incident wave, and $E_\parallel^0(\omega)$ is the transmitted parallel field through the substrate only. Likewise, let the perpendicular transmittance $T_\perp(\omega) = E_\perp(\omega)/E_\parallel^0(\omega)$ where $E_\perp(\omega)$ is the perpendicular component of the transmitted field. Then it can be shown that for $\sigma_{xy}/\sigma_{xx} \ll 1$:

$$\begin{aligned}\sigma_{xx} &= [1/T_\parallel - 1]/\alpha \\ \sigma_{xy} &= -[T_\perp/T_\parallel^2]/\alpha\end{aligned} \quad (2)$$

where $\alpha = \mu_0 c/(n_s + 1)$ and $n_s$ is the substrate index of refraction.

## II. Experiment

Single layer graphene samples were grown on copper foil by chemical vapor deposition (CVD) following Li *et. al.* [25]. After growth the graphene films were transferred to 1cm$^2$ sapphire substrates using the conventional wet chemical approach. Electrical resistivity and Hall Effect measurements of as-transferred films showed the samples to be *p*-type with carrier concentrations in the range $p = 2\cdot 10^{12}$cm$^{-2}$ - $2\cdot 10^{13}$cm$^{-2}$, and DC mobilities $\mu$ = 500-2000 cm$^2$/Vs. Infrared transmission measurements were performed using a Thermo-Nicolet IS-50 FTIR with near-, mid-, and far-infrared optics. Far-infrared transmission measurements used a 4.2 K Si bolometer (IR Labs) as a detector.

Our time-resolved THz system is based on a FemtoLasers XL500 chirped pulse oscillator (800nm center wavelength, 5MHz repetition rate, 0.5μJ pulse energy, 50fs pulse-width). The laser pulses are divided into Pump, THz Generation and Probe beams, each beam incorporating an independent delay stage. The Generation beam is incident on a photoconductive switch to generate horizontally polarized pulses of THz radiation. The THz pulses are collected and focused onto the sample, mounted in an Oxford Spectromag magneto-optical cryostat with THz transparent quartz windows. The magnetic field is parallel to the beam direction. The transmitted THz beam is focused onto a 1mm-thick ZnTe crystal, where the THz field amplitude is measured by the Probe beam through the electro-optic effect. Sweeping the THz beam delay allows measurement of the THz pulse waveform. Wire-grid polarizers are placed before and after the sample (Fig. 1). The first polarizer is set to pass horizontally polarized radiation, and the second polarizer and the electro-optic detector are rotated together to detect either the horizontally or vertically polarized components of the transmitted radiation. The photoconductive emitter voltage is modulated and the signal is recovered with a lock-in amplifier. For time-resolved measurements, the Pump pulse excites the samples at an incident photon flux $\phi \sim 1 \cdot 10^{13} cm^{-2}$. The Pump beam is chopped and the signal is recovered using two lock-in amplifiers. Sweeping the Pump delay allows measurement of the change in THz transmission as a function of time following photoexcitation.

For linear transmission measurements, we measured the time dependent-electric field of THz pulses transmitted through a sample and through a reference sample, in both polarizations. Fourier Transforms of these waveforms yielded the sample and reference electric fields versus frequency. From these we determined the parallel and perpendicular transmittance, $T_\parallel(\omega)$ and $T_\perp(\omega)$ and the equilibrium conductivity tensor components $\sigma_{xx}(\omega)$ and $\sigma_{xy}(\omega)$.

We tested this novel technique on well-characterized GaAs and InP samples with carrier mobilities similar to CVD graphene. Figure 2 shows linear magneto-transmission measurements of an *n*-type GaAs epitaxial layer on a semi-insulating GaAs substrate with nominal doping of the 7μm thick epi-layer of $n = 1.0 \cdot 10^{17} \text{cm}^{-3}$. We found that the transmitted THz waveform shape varied only weakly with magnetic field, so that it was sufficient to track the peak amplitude of the waveform versus magnetic field. The sheet carrier density and mobility obtained agreed within 20% with values obtained from a Drude fit to the frequency dependent transmission.

Time-resolved measurements were carried out by additionally measuring the change in the parallel and perpendicular transmission after photoexcitation, $\Delta T_\parallel$ and $\Delta T_\perp$. Test measurements were carried out on semi-insulating InP:Fe, in which the pump-induced photoconductivity is dominated by the electrons (Fig. 3). In this case, the photoexcited layer is treated as the sample, the semiconductor substrate can be treated as the reference, and analysis yields the carrier density and scattering rate of the photoexcited layer (Fig. 4). The THz probe pulse interrogated the sample 10ps after photoexcitation. The measured photocarrier density was within 10% of the absorbed photon flux and the photocarrier mobility was within 20% of the mobility determined by frequency-dependent time resolved THz spectroscopy measured under similar conditions.

## III.    RESULTS

Linear magneto-transmission measurements were carried out on single-layer graphene samples to determine the conductivity tensor elements $\sigma_{xx}$ and $\sigma_{xy}$ (Fig. 5). The sample was in a helium gas ambient at temperature T = 240K. For this sample, the conductivity was found to be nearly independent of frequency over the THz range investigated by this experiment, indicating that the scattering rate was much larger than the maximum frequency accessible in our THz measurements. In this case, the amplitude of the THz field could be tracked by measuring the peak amplitude of the THz waveform versus magnetic field. As in the *n*-GaAs sample, we find $\sigma_{xx}$ to be nearly symmetric in the magnetic field, while $\sigma_{xy}$ is antisymmetric. The ratio $\sigma_{xx}/\sigma_{xy}$ is approximately linear in the region near $B = 0$. The positive slope shows that the sample is *p*-type and the magnitude of the slope gives the hole mobility $\mu = 2400 \text{cm}^2/\text{Vs}$. From the $B = 0$ sheet conductivity $\sigma_{xx} = 7.7 \cdot 10^{-4}\ 1/\Omega$, we obtain a carrier concentration $p = 2.0 \cdot 10^{12} \text{cm}^{-2}$ and average scattering time $\tau = 37\text{fs}$. We observe a significant discrepancy (~50%) between these values (taken in He atmosphere) and values obtained from FTIR measurements taken at room temperature in air, and ascribe this difference the strong environmental sensitivity[26] of graphene.

Figure 6 shows the change in conductivity due to photoexcitation as a function of delay. We find $\sigma_{xx}$ *decreases* after photoexcitation (negative photoconductivity) and recovers with 1.2ps time-constant, and that $\Delta\sigma_{xx}$ is approximately symmetric with magnetic field. The change $\Delta\sigma_{xy}$ is approximately antisymmetric in magnetic field and decays with a similar time constant. We have also measured the conductivity change as a function of magnetic field, at fixed delay. Figure 7 shows the fractional conductivity change measured 0.5ps after photoexcitation. The fractional change $\Delta\sigma_{xx}/\sigma_{xx}$ is a nearly field independent, while $\Delta\sigma_{xy}/\sigma_{xx}$ is approximately proportional to magnetic field.

## IV.     DISCUSSION

To analyze our results we have developed a model of transient photoconductivity in graphene based on a theory by Rana, *et. al.*,[27] The model, described in detail elsewhere[28], assumes that photoexcited electrons and holes thermalize within ~200fs of photoexcitation. At later times the electron and hole distribution functions are described by a single carrier temperature and chemical potential. The hot carriers exchange energy with the lattice by optical phonon emission and absorption, and carrier-phonon system cools as the optical phonons decay to acoustic phonons. The model tracks the total kinetic energy of the free carriers, carrier temperature, electron and hole carrier densities, phonon populations, and carrier scattering rates. At each time step the conductivity is determined from the energy-dependent carrier scattering rates and electron and hole distribution functions:

$$\sigma_{xx}^e = \int_{-\infty}^{\infty} \frac{\sigma_0^e(\varepsilon)}{1+\omega_c^2(\varepsilon)\tau_e^2(\varepsilon)} d\varepsilon, \qquad \sigma_{xy}^e = \int_{-\infty}^{\infty} \sigma_0^e(\varepsilon) \frac{\omega_c(\varepsilon)\tau_e(\varepsilon)}{1+\omega_c^2(\varepsilon)\tau_e^2(\varepsilon)} d\varepsilon, \qquad (3)$$

where[29] $\sigma_0^e(\varepsilon) = -\frac{1}{2}e^2 v_f^2 g_e(\varepsilon)\tau_e(\varepsilon)(\partial f_e/\partial \varepsilon)$. Here $\tau(\varepsilon)$ is the energy-dependent scattering time, $f_e$ is the Fermi function and $g_e(\varepsilon)$ is the density of states. The orbital cyclotron frequency in graphene[30] is $\omega_c = \left(qv_F/\hbar\sqrt{\pi p}\right)B$ where $q$ is the carrier charge. A similar expression holds for holes, and $\sigma_{ij} = \sigma_{ij}^h + \sigma_{ij}^e$.

This model was fit to our time-resolved conductivity measurements (Fig. 6), taking the initial absorbed energy as an adjustable parameter, and taking the equilibrium carrier concentration and scattering rate from our linear THz Faraday Rotation measurements. The same parameters were used to fit the field-dependent measurements at fixed delay. In Fig. 7, the solid line shows the model fit, which is in good agreement with our measurements. The dashed line shows the hole conductivity only, calculated for the same set of parameters. According to the model, the photoconductivity of this strongly *p*-type sample is controlled by the majority carriers.

Interestingly, the detailed model described above can be seen to be nearly equivalent to a simpler effective model. Treating the photoexcited graphene within the Drude model, with a single effective carrier scattering rate, and ignoring the contribution of the minority carriers to the conductivity, we can extend the analysis above to the case of graphene using the zero-field conductivity $\sigma_0 = \left(e^2 v_F \tau/\hbar\right)\sqrt{p/\pi}$. Then, for small perturbations in the carrier concentration and the scattering rate we find

$$\frac{\Delta\sigma_{xx}}{\sigma_{xx}} = \left(\frac{\Delta p}{2p} + \frac{\Delta\tau}{\tau}\right); \quad \text{and} \quad \frac{\Delta\sigma_{xy}}{\sigma_{xx}} = 2\omega_c \Delta\tau. \qquad (4)$$

The ratio $\Delta\sigma_{xx}/\sigma_{xx}$ should be independent of field, while the ratio $\Delta\sigma_{xy}/\sigma_{xx}$ should be proportional to magnetic field. The experimental results closely match the prediction of this simplified model, and our results yield a decrease in effective scattering time, $\Delta \tau = -(0.068 \pm 0.001)\tau$ and an increase in carrier concentration $\Delta p = (0.019 \pm 0.002)p$. Within the Drude approximation the conductivity change is dominated by the reduction in the effective carrier scattering time.

## V.  CONCLUSION

We have shown that THz Faraday Rotation measurements can independently determine the carrier density and effective scattering rate in thin conducting films, and that time-resolved measurements can independently determine the change in carrier density and effective scattering rate in photoconductive materials.

We have applied this technique to p-type CVD graphene samples that exhibit negative photoconductivity. Our results can be fit with a numerical model based on energy dependent electron-phonon scattering. However, they also can be fit by a simplified model in which majority carriers scatter at single effective rate. In our measurement we find a 7% decrease in the effective scattering lifetime and a 2% increase in carrier concentration 0.5ps after photoexcitation with an incident photon flux of $1 \cdot 10^{13} \text{cm}^{-2}$. Our results are consistent with models in which the carrier distribution after photoexcitation of graphene is thermal with a single chemical potential and carrier temperature in the valance and conduction bands within 0.5ps of photoexcitation.

## VI.  ACKNOWLEDGEMENTS

The authors acknowledge the contribution of Mr. Yilikal Ayino in building the THz magnetospectroscopy system. This material is based upon work supported by the National Science Foundation under Grant No. DMR-1006065. Research at NRL was supported by Base Programs funded through the Office of Naval Research.

FIGURES

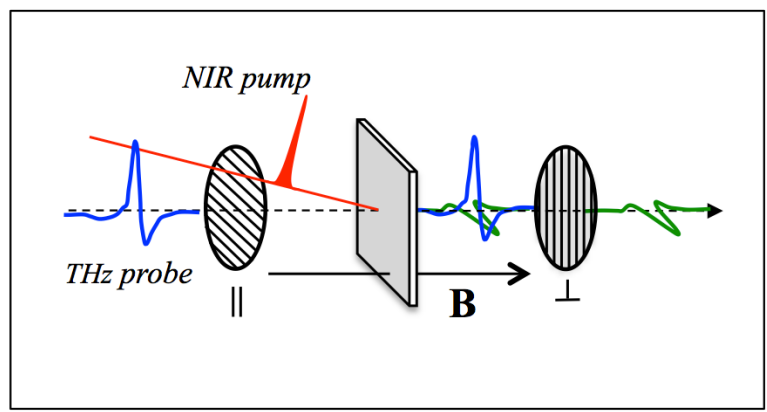

Figure 1. Geometry of the THz Faraday Rotation measurements. Linearly polarized THz pulses are normally incident on a sample through a wire-grid polarizer. A magnetic field normal to the sample surface rotates the polarization of the transmitted pulse and a second wire-grid polarizer is used to analyze the polarization. A NIR pump pulse is incident on the sample for time-resolved measurements.

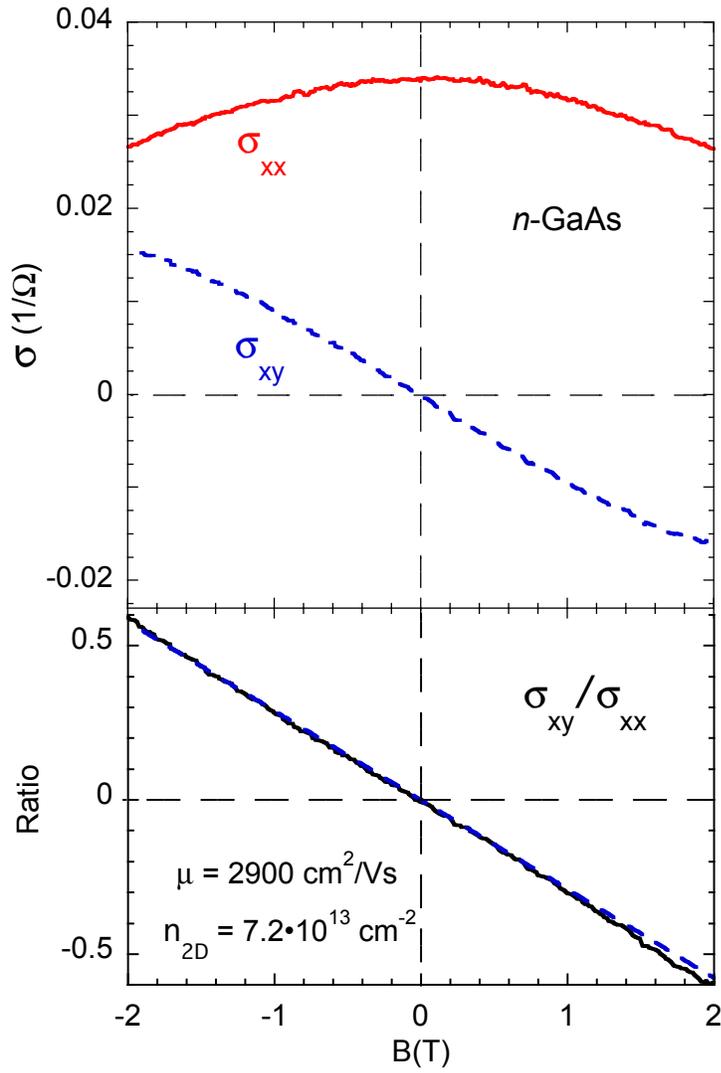

Figure 2 (a) Longitudinal ($\sigma_{xx}$) and transverse ($\sigma_{xy}$) conductivity of an n-GaAs epitaxial layer versus magnetic field extracted from polarization-dependent THz measurements. (b) The ratio $\sigma_{xy}/\sigma_{xx} = \omega_c \tau = \mu B$ directly yields the electron mobility.

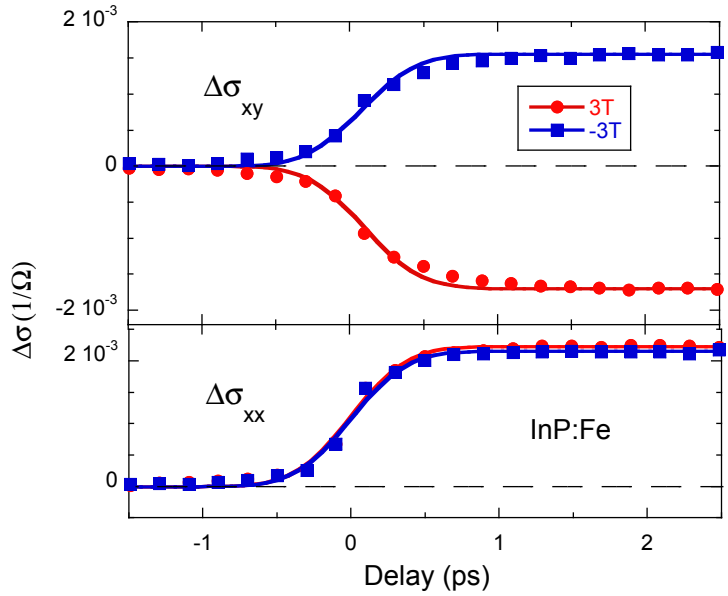

Figure 3  The pump-induced change in the longitudinal and transverse conductivity versus delay in semi-insulating InP:Fe at $B=\pm 3T$. $\Delta\sigma_{xx}$ is symmetric in the magnetic field, while $\Delta\sigma_{xy}$ is antisymmetric.

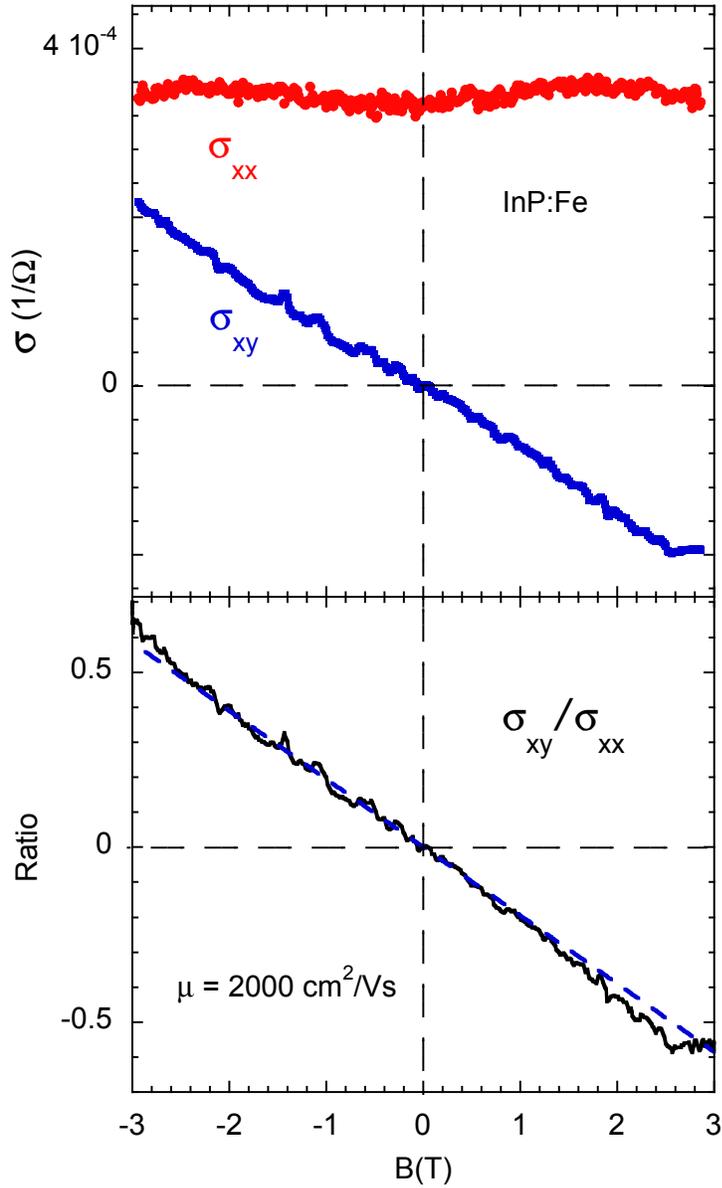

Fig. 4    Magnetic field dependence of the pump-induced conductivity of InP:Fe. In this semi-insulating material the photoconductivity is dominated by the electrons, so that the ratio $\sigma_{xy}/\sigma_{xx} = \omega_c \tau = \mu B$ directly yields the photoelectron mobility.

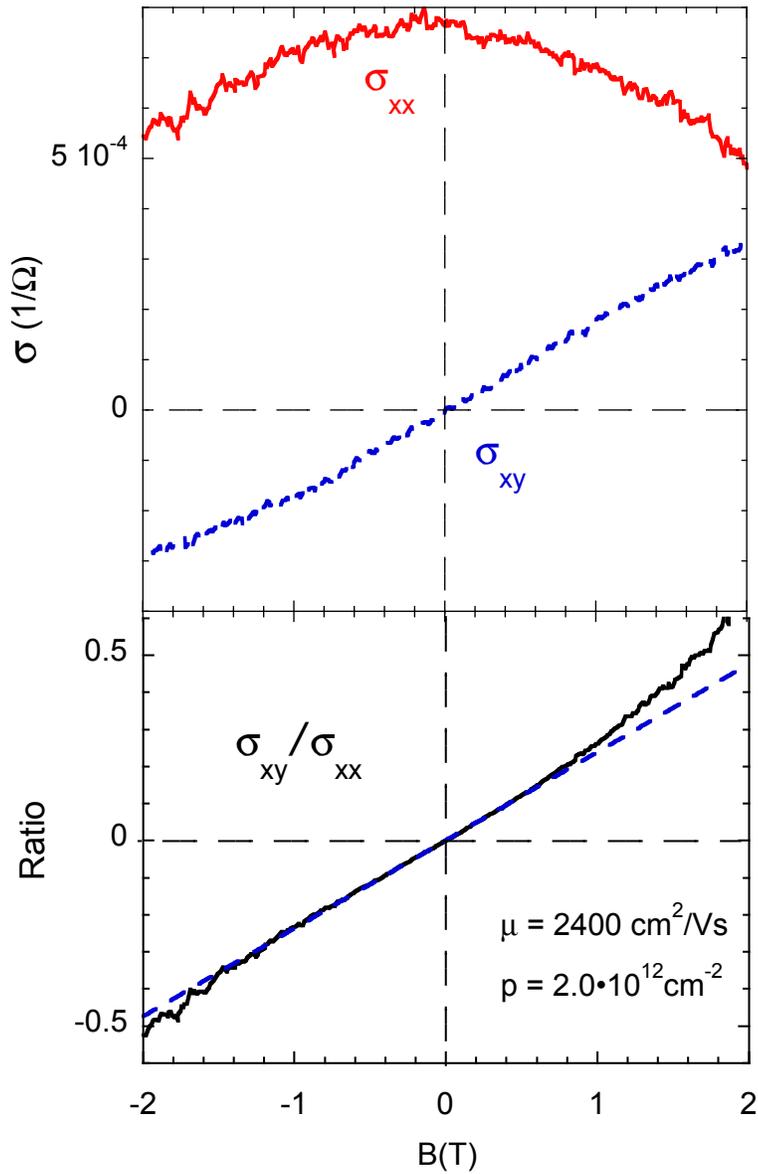

Figure 5 (a) Longitudinal ($\sigma_{xx}$) and transverse ($\sigma_{xy}$) conductivity of *p*-type CVD graphene versus magnetic field extracted from polarization-dependent THz measurements. (b) The ratio $\sigma_{xy}/\sigma_{xx} = \omega_c\tau = \mu B$ yields the hole mobility.

*Heyman et. al.,* THz Faraday Rotation 14

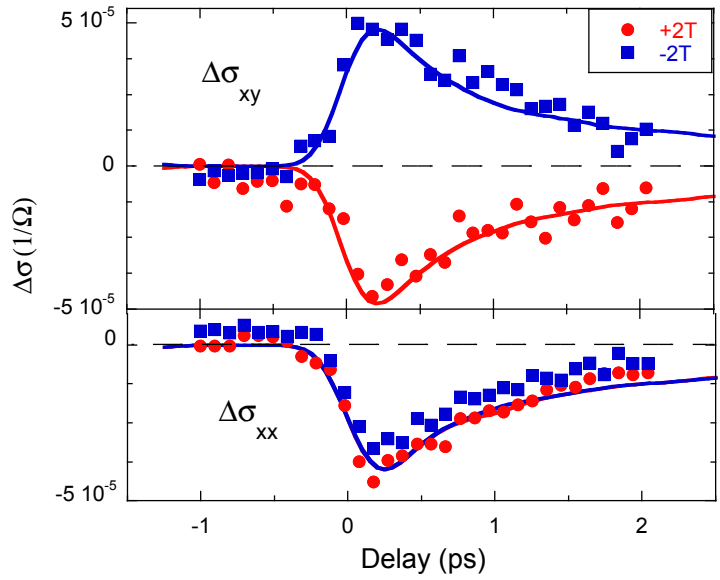

Fig 6. Change in the longitudinal and transverse conductivity due to photoexcitation in *p*-type CVD graphene at $B=\pm 2T$. $\Delta\sigma_{xx}$ is symmetric in the magnetic field, while $\Delta\sigma_{xy}$ is antisymmetric. The solid lines are model fits.

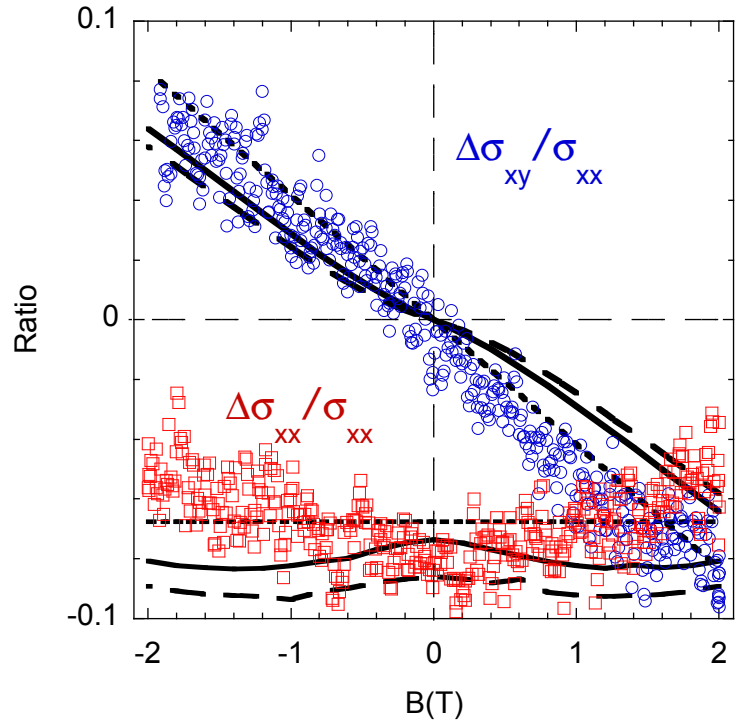

Fig. 7  The pump-induced change in longitudinal and transverse conductivity of the *p*-type CVD graphene sample, divided by the equilibrium longitudinal conductivity.  The solid line is a fit to our full numerical model.  The dashed line is a numerical fit including the photo-excited holes only.  The dotted line is an effective model with a single carrier type and single scattering rate.